\documentclass[12pt]{iopart}
\usepackage{graphicx}
\usepackage{colordvi,color}

\newcommand{\beq}{\begin{equation}}
\newcommand{\eeq}{\end{equation}}
\newcommand{\bea}{\begin{eqnarray}}
\newcommand{\eea}{\end{eqnarray}}

\newcommand{\Dh}{\Delta h}

\newcommand{\figone}{%
\begin{figure}[htbp]
   \centering
   \includegraphics[width=9cm,height=4.5cm,clip]{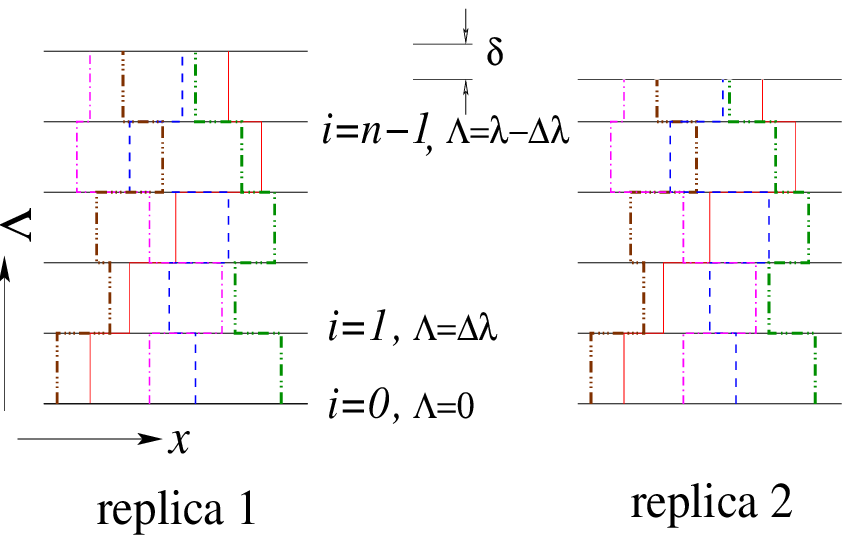}
   \caption{%
     Schematic representation of two replicas of same paths, each
     starting from $\Lambda=0$ and ending at $\Lambda=\lambda$ in
     replica $1$ and at $\Lambda=\lambda-\delta$ in replica $2$. Label
     $i$ denotes the step number as $\Lambda$ is changed in steps of
     $\lambda/n$.  Lines of different styles (dashed, dotted etc)
     represent different realizations of paths starting from different
     values of $x$.  The vertical portion of a path is an
     instantaneous process (no change in $x$) and the horizontal part is
      under interaction with the surrounding  ($x$ evolves
       at a constant $\Lambda$). Identically shaded lines in the two
     replicas have the same evolution. }
   \label{fig:fig1}
 \end{figure}
}
\newcommand{\figtwo}{%
\begin{figure}[htbp]
   \centering
   \includegraphics[width=12cm,clip]{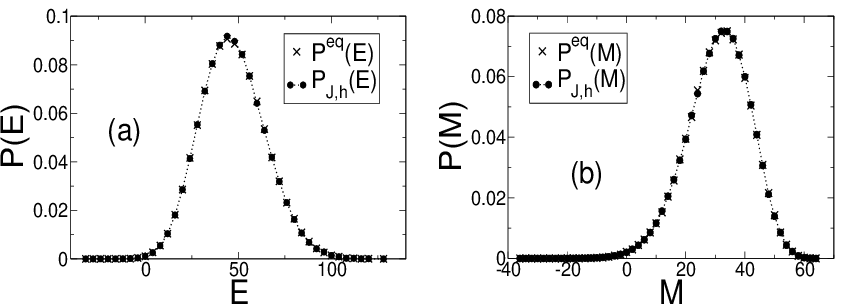}
   \caption{%
     Plot of the weighted distribution (a) $P_{J,h}(E)$ vs. $E$ and
     (b) $P_{J,h}(M)$ vs. $M$ (dotted line with circles) for varying
     $J$ and $h$ with $n=20$ and equilibrium distributions $P_{eq}(E)$
     and $P_{eq}(M)$ (crosses) with $J\!=\!1$, $h=1$ and $\beta=0.2$
     for a $8\times 8$ lattice, showing that $P_{J,h}(E)=P_{eq}(E)$
     and $P_{J,h}(M)=P_{eq}(M)$.}
   \label{fig:fig2}
 \end{figure}
}
\newcommand{\figthree}{%
\begin{figure}[htbp]
   \centering
   \includegraphics[width=12cm,width=6cm,clip]{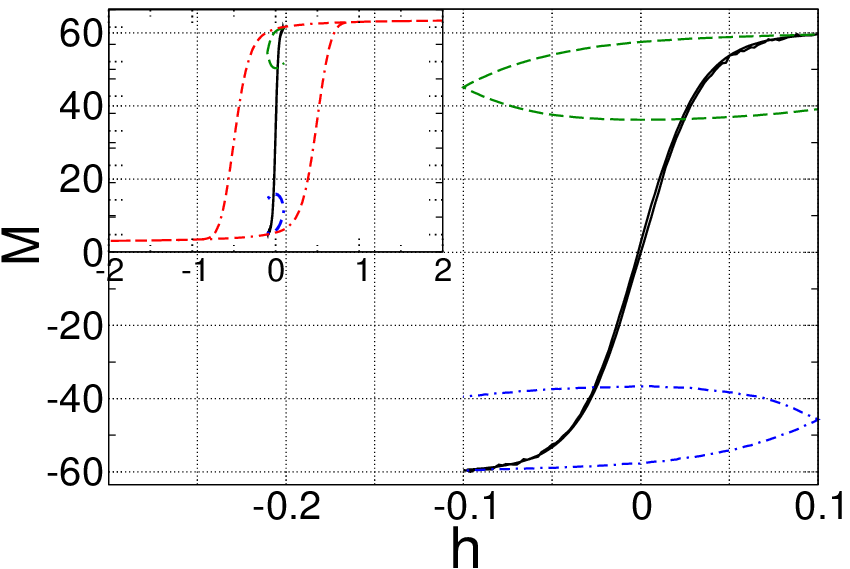}
   \caption{%
     Plot of weighted average $M(h)$ vs. $h$ (black solid line) and
     hysteresis loop, simple averaged $M$ vs. $h$ (green dashed and
     blue dash-dotted lines) for a $8\times 8$ lattice. The magnetic
     field $h$ varies from $-0.2$ to $+0.2$ in $100$ steps. Inset
     shows the hysteresis loop for the small (green and blues lines)
     field with respect to the large field (red double dash-dotted
     line) and the weight averaged magnetization for small field.  }
   \label{fig:fig3}
 \end{figure}
}
\newcommand{\figfour}{%
\begin{figure}[htbp]
   \centering
   \includegraphics[clip]{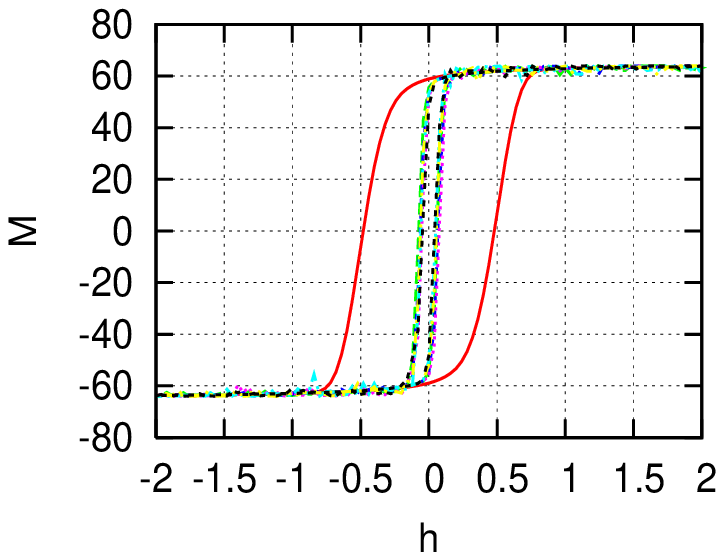}
   \caption{%
     Plot of weighted average $M(h)$ vs. $h$ (dashed lines) and
     hysteresis loop, simple averaged $M$ vs. The magnetic field $h$
     (red solid line) for a $8\times 8$ lattice.  $h$ varies from $-2$
     to $+2$ in $100$ steps.  }
   \label{fig:fig4}
 \end{figure}
}
\newcommand{\figfive}{%
\begin{figure}[htbp]
   \centering
   \includegraphics[width=12cm,clip]{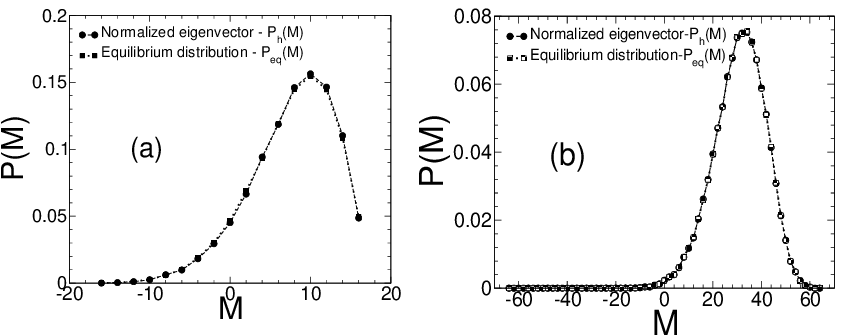}
   \caption{%
     Plot of the equilibrium distribution $P_{eq}(M)$ vs. $M$ (boxes
     with dotted line) and normalized principal eigen-vector $P_h(M)$
     (dashed line with circles) with $J=1$, $h=1$, $\beta=0.2$ and
     $n=1000$ for (a) $4\times 4$ lattice and (b) $8\times 8$ lattice,
     showing that $P_h(M)=P_{eq}(M)$, i.e., eigenfunction is indeed an
     equilibrium distribution.  }
   \label{fig:fig5}
 \end{figure}
}
\usepackage{iopams}
\begin{document}

\title{Thermodynamics as a  nonequilibrium path integral}

\author{Poulomi Sadhukhan and Somendra M. Bhattacharjee}

\address{Institute of Physics, Bhubaneswar 751 005, India}
\ead{poulomi@iopb.res.in, somen@iopb.res.in}
\begin{abstract}
  Thermodynamics is a well developed tool to study systems in
  equilibrium but no such general framework is available for
  nonequilibrium processes. Only hope for a quantitative description
  is to fall back upon the equilibrium language as often done in
  biology. This gap is bridged by the work theorem. By using this
  theorem we show that the Barkhausen-type nonequilibrium noise in a
  process, repeated many times, can be combined to construct a special
  matrix ${\cal S}$ whose principal eigenvector provides the
  equilibrium distribution. For an interacting system ${\cal S}$, and
  hence the equilibrium distribution, can be obtained from the free
  case without any requirement of equilibrium.
\end{abstract}

\pacs{05.70.Ln 05.20.-y  82.20.Wt 87.10.-e}
\submitto{\JPA}
\maketitle

\section{Introduction}
A system in thermodynamic equilibrium has no memory of its past.
Consequently there is no leading role for time in the ensemble based
statistical mechanics except the subservient one to maintain
equilibrium among the internal degrees of freedom and with external
sources. This wisdom gets exploited in the dynamics based algorithms
like Monte Carlo, molecular dynamics, stochastic quantization, to name
a few, to attain equilibrium from any arbitrary state albeit in
infinite time.  Even a thermodynamic process involving changes in
parameters is an infinite sequence of equilibrium states, and is
therefore infinitely slow. A finite duration process, not destined to
equilibrate at every instant of time, maintains a memory of the
initial conditions or a short time correlation of states.  The biased
sampling of the phase space keeps these processes outside the realm of
statistical mechanics and thermodynamics.  In this
equilibrium-nonequilibrium dichotomy, a work
theorem\cite{boch,Jar,cro,hum,coh} attempts to bridge the gap by
providing a scheme for getting the thermodynamic free energy
difference from a properly weighted nonequilibrium path
integral\cite{cro,hum}.

We show in this paper that purely nonequilibrium measurements of work
gives an operator ${\cal S}$, defined on the phase or configuration
space, whose normalized principal right eigenvector is the equilibrium
probability distribution.
Our result is valid for any number of parameters including temperature
and interaction. With this extension we can get the equilibrium
distribution by constructing a matrix $\cal S$ connecting any two
allowed states of the system without any reference to equilibrium
anywhere, thereby completely blurring the boundary between equilibrium
and nonequilibrium. This finds direct application in
out-of-equilibrium phenomena like hysteresis.

Barkhausen noise is an example of nonequilibrium response of a
ferromagnet as the magnetic field is changed at a given
rate\cite{bar,hyst}.  By measuring the voltage induced in a secondary
coil as the current in the primary coil wound around a ferromagnet is
changed, one gets the time variation of the magnetization.  The noisy
signal one gets is not unique but stochastic in nature, reflecting the
fluctuating microscopic response to the external field.  Such signals
have been analyzed in the past to extract information like avalanche
statistics, material characteristics etc.  Our results find a
different use of the Barkhausen noise to construct the ${\cal S}$
matrix.  Similar constructions  for other cases like
protein or DNA dynamics in vivo, pulling of polymers in
single-molecule experiments, etc,  call for new class of
experiments to monitor the noise signals during these
events.

This paper is organized as follows: In Sec. \ref{sec:work}, we
recapitulate the work theorem, introduce the paths and discuss the
connection between the work theorem and the histogram transformation
of equilibrium statistical mechanics. In Sec.  \ref{sec:prob} we give
a simple and general, dynamics independent proof of the relation
between the equilibrium probability distribution and the work done in
nonequilibrium paths.  This relation in some form is already known
\cite{hum,cro} but our derivation allows us in generalizing the result
to other cases involving temperature, interactions, etc.  Sec.
\ref{sec:opS} deals with the main result of this paper. There we prove
the eigenvalue equation for ${\cal S}$.  A few examples are also given
there.  How to get the operator ${\cal S}$ directly from experimental
measurements of Barkhausen noise is also discussed here.  Numerical
verifications of some of the results are presented in Sec.
\ref{sec:num} by taking the $2D$ Ising model as an example. We
summarize in Sec. \ref{sec:summ}.
\section{Work theorem and path integral}\label{sec:work}
\subsection{Work theorem}
Consider a classical system described by a Hamiltonian $H(\Lambda,x)$
where ${\Lambda}$ is an external field that couples to its conjugate, a
microscopically defined quantity, $x$. The thermodynamic state is
specified by temperature $T$ and field $\Lambda$. Let us start with
the system at $\Lambda=0$ in thermal equilibrium at temperature $T$.
External field $\Lambda$ is changed in some given way from $0$ to a
final value $\lambda$ in a finite time $\tau$ or in a finite number of
steps $n$, letting the system evolve in contact with the heat
reservoir. No attempt is made to ensure equilibrium during the
process. The variation of $x$ along the nonequilibrium path ($x(t)$ vs
$t$) and the instantaneous final (boundary of the path) value of $x$,
$x_{\rm b}$, when the field reaches $\lambda$, are noted. The work
done along a nonequilibrium path by the external source (as in ref.\cite{Jar}) is
\begin{equation}
 \label{eq:1}
W=\int_0^\tau \frac{ \partial H}{\partial\Lambda}\frac{d\Lambda}{dt}\,dt ,
\end{equation}
in time $\tau$, and it varies from path to path.
The difference between two definitions of work in the context of work theorem, one used in ref. \cite{boch} and the other in ref.\cite{Jar}, is discussed in ref. \cite{hor}.
For the sake of notational simplicity
we choose,
\begin{equation}
  \label{eq:22}
H=H_0+H_1(\Lambda,x)=H_0-\Lambda\ x,
\end{equation}
where $H_0$ is the energy for $\Lambda=0$. There is not much loss of
generality in choosing the form of Eq. \ref{eq:22} because $\Lambda$
and $x$ refer to any pair of conjugate variables so that $x$ itself
need not be a linear function of the internal coordinates.  As an
example, in an interacting spin problem in a magnetic field $h$
($\equiv\! \Lambda$), $H=H_0-h\sum_k s_k$ where $s_k$ is the spin
variable at a site denoted by $k$, with $x=\sum_k s_k$.  Often
$\Lambda$ can be taken as the switching parameter to turn on a
perturbation or interaction in a Hamiltonian $H=H_0-H'$ with
$H_{\Lambda}=   H_0 -\Lambda H'$.

The work theorem\cite{boch,Jar} provides the equilibrium free energy
difference $\Delta F$ between the two states with $\Lambda=0$ and
$\Lambda=\lambda$, both at inverse temperature $\beta=1/k_BT$ ($k_B$
is the Boltzmann constant), from the nonequilibrium work done as
\beq
 \label{eq:2}
\Delta F= - \frac{1}{\beta} \ln \langle e^{-\beta W}\rangle ,
\eeq
where $\langle...\rangle$ denotes the average over all possible paths.
\subsection{Paths: equilibrium and nonequilibrium}\label{sec:paths:-equil}
We are using here a description of a state by the intensive parameters
which actually characterize the surroundings.  In equilibrium any
system is expected to have the values of the intensive parameters same
as that of the environment. A change in any of the parameters, say
$\Lambda$, from $\lambda_0$ to $\lambda$, would require heat and/or energy transfer.   The work done on
or by the system is determined by the change in the free energies,
independent of the path of variation of the intensive parameters.  This is
expressed as
\begin{equation}
\label{eq:25}
  \Delta F= W_{\rm eq} =-\int_{\lambda_0}^{\lambda} x_{\rm eq}(\Lambda)\  d\Lambda,
\end{equation}
where $\Delta F= F(\beta,\lambda)-F(\beta,\lambda_0)$.
Here $ x_{\rm eq}(\Lambda) =\int x\, P_{\Lambda}(x) dx$ is the equilibrium
average at the instantaneous values of the intensive parameters and
$P_{\Lambda}(x)$ is the corresponding equilibrium probability
distribution of $x$.
This follows from the identification of the equilibrium value of $x$
as $ x_{\rm eq}=-\partial F/\partial\Lambda$, in contrast to the
conjugate ensemble definition $\Lambda=\partial {\cal F}/\partial x$
where ${\cal F}(\beta,x)$ is the fixed-$x$ ensemble free energy.

For convenience, let us discretize the integrals.  For example, for
$\Lambda \in [\lambda_0,\lambda]$, we have a sequence $(\Lambda_0,
\Lambda_1,...\Lambda_n=\lambda)$ and the continuum is recovered by
taking the usual limit of $n\to\infty$ with
$\max\{\Delta\Lambda_i=\Lambda_{i+1}-\Lambda_i\}\to 0.$ The work done
can be rewritten as
\begin{equation}
\label{eq:24}
W_{eq}=-\sum_{i=0}^{n-1}  \Delta\Lambda_i\ \left\lbrace \sum_x
P_{\Lambda_i}(x) x \right\rbrace .
\end{equation}
By interchanging the sums over $x$ and $\Lambda$, we define (i) a
sequence  $\{x_i|i=0,...n\}$ as instantaneous values, and (ii) a
sequence-dependent work done as $W=\sum_i x_i\:
\Delta\Lambda_i$, to reinterpret Eq. (\ref{eq:24}) as an average over these
$x_i$'s.  Therefore,
\begin{equation}
\label{eq:23}
W_{eq}=-\sum_{\{x_i\}} {\cal P}\{x_i\} \sum_i x_i\, \Delta\Lambda_i,
\end{equation}
where $ {\cal P}\{x_i\} = \prod_i P_{\Lambda_i}(x_i) $ is the joint
probability of getting the particular $\{x_i\}$ sequence, because, for a
thermodynamic process, there is no memory.  Going over to the
continuum limit, the thermodynamic process of varying $\Lambda$ is now
seen as equivalent to choosing a path in the configuration space and
re-weight the paths according to the probability of its occurrence in
the $\Lambda$-ensemble.  The relation between the free energy change
and work, Eq. (\ref{eq:25}), now gets a path integral meaning where
the process takes the system over the microstates and one averages the
work over individual paths.

This thermodynamic connection is valid only in equilibrium.  The work
theorem generalizes this idea  by replacing  $ {\cal P}\{x_i\}$ by the
nonequilibrium probability of getting a path and asserting
\begin{equation}\label{eq:26}
e^{-\beta \Delta F}\equiv \frac{Z_\lambda}{Z_0}=\int {\cal D X}\  e^{-\beta W},
\end{equation}
where $\int {\cal DX}$ stands for the normalized sum over paths, i.e.,
sum over intermediate $x$'s with appropriate probabilities.
\subsection{Histogram transformation and infinitely fast process}\label{sec:histo}
There is a fundamental transformation rule obeyed by the partition
function, often used in numerical simulations as the histogram
method\cite{fal}. This transformation connects the equilibrium
probability distributions at two parameter values, $\Lambda=\lambda_0$
and $\Lambda=\lambda$ as
\beq
 \label{eq:5}
 P_{\lambda}(x)=\frac{P_{\lambda_0} (x)\ e^{\beta
     ({\lambda}-{\lambda}_0)x}}{\sum_x P_{{\lambda}_0}(x)\ e^{\beta
     ({\lambda}-{\lambda}_0)x}},
\eeq
where the sum in the denominator is over the allowed values of
$x$. The denominator of the right hand side of Eq.\ref{eq:5} is
$Z_{\lambda}/Z_{\lambda_0}$ where $Z_{\lambda}$ is the partition
function at inverse temperature $\beta$,
\begin{equation}
  \label{eq:27}
  Z_{\Lambda}~=~\sum_{states}\,e^{-\beta  H_0}\,e^{\beta\Lambda x}.
\end{equation}
From Eq. \ref{eq:1},
$(\lambda-\lambda_0)x$ can be taken as the work done in an
instantaneous process that changes $\Lambda$ from $\lambda_0$ to
$\lambda$ without changing $x$. The probability of getting $x$ for
equilibrium at $\lambda_0$ is $P_{\lambda_0}(x)$ and therefore the sum
in the
denominator of Eq. \ref{eq:5} is the path integral of
Eq. \ref{eq:26}, because $x$ does not change.
This gives the work theorem.
\section{Equilibrium probability distribution}\label{sec:prob}
We in this section use the discrete version of the process to re-derive
the equilibrium probability distribution from the work theorem in a
general and dynamics independent way.
For the kind of
nonequilibrium processes mentioned in Sec. \ref{sec:paths:-equil} the
equilibrium probability distribution of $x$ at a parameter value
$\lambda$ can be obtained from a weighted path integral\cite{cro,hum}
\begin{equation}
  \label{eq:3}
   P_\lambda(x) =
          \frac{\int {\cal DX}\ e^{-\beta W}\ \delta (x_{\rm b}-x)}{\int
            {\cal DX} \; e^{-\beta W}} ,
\end{equation}
where $x_{\rm b}$ is the instantaneous boundary value at the end of
the path, and  the denominator is same as r.h.s. of Eq. \ref{eq:26}.
This is in the form of a path integral where the paths are weighted by
a Boltzmann-like factor $\exp(-\beta W)$.  The same was established
previously in specific cases like, the Master equation
approach\cite{Jar}, the Feynman-Kac formula\cite{hum} and Monte Carlo
dynamics\cite{cro}.

The equilibrium average  $x_{\rm eq}$ is defined as
\begin{equation}
 \label{eq:A3}
     x_{\rm eq}= \frac{1}{\beta}\frac{\partial \, }{\partial \Lambda}
  \ln Z_\Lambda =
     \lim_{\delta\rightarrow 0}\, \left(\beta
       \frac{Z_\Lambda}{Z_0}\right)^{-1} \;\frac{1}{\delta}\  \left(
  \frac{Z_\Lambda}{Z_0}-\frac{Z_{\Lambda-\delta}}{Z_0}\right),
\end{equation}
where work theorem is to be used for the partition functions.

The  system starts in equilibrium at temperature $T$ and
$\Lambda=0$, and then  $\Lambda$ is built up at constant $T$  as a sequence
of infinitely fast jump of
$\Delta \lambda=\lambda/n$, each jump followed by a finite time
evolution in contact with the heat bath.
Consider now two $n$-step processes,  one process with final
field $\lambda$ and another one with $\lambda-\delta$
($\delta\rightarrow 0$ at the end).
In fact, the second process is just a copy (replica) of
the first one in every respect except at the last stage
(Fig. \ref{fig:fig1}).
For the last jump, the change in $\Lambda$ for replica 1 is
$\Delta\lambda$ while for replica 2 it is $\Delta\lambda -\delta$.

\figone

A path is specified or defined by the sequence
$\{x_i\mid i=0,... n-1\}$. The changes in $x_i$ at any step is
because of internal dynamics or exchange of heat with the external
reservoirs. We do not need to let the system evolve once the field
reaches the final desired value.   Therefore, the sequence
$\{x_i\mid i=0 ...{n-1}\}$ is the same for both the replicas.
The work done $W_1, W_2$ along an
$n$-step nonequilibrium path for replicas $1,2$ are related via
\beq
 \label{eq:A5}
W_2=W_1+\delta\, x_{n-1},
\eeq
with $W_1$ is of the form given above Eq. (\ref{eq:23}).
The work theorem of  Eq. (\ref{eq:26})
when used in Eq. \ref{eq:A3} yields
\begin{eqnarray}
  x_{\rm eq}
&=& \lim_{\delta \rightarrow 0}\
      \frac{1}{\beta \sum_{\rm paths}e^{\beta
\sum_{i=0}^{n-1}\Delta \Lambda_i x_i}}
       \   \frac{\sum_{\rm paths} e^{\beta \sum_{i=0}^{n-1}\Delta \Lambda_i  x_i}
            \left( 1-e^{-\beta \delta x_{n-1}}\right) }{\delta}
              \nonumber\\
&=& \frac{\int {\cal DX}\ x_{\rm b}\ e^{-\beta W}}{\int {\cal DX}\
  e^{-\beta W}} , \ \ (x_b\equiv x_{n-1}).\label{eq:A8}
\end{eqnarray}
This shows that the equilibrium average can be expressed in terms of
the boundary value with proper weightage of the paths.
The above proof can be generalized to any moments of $x$.

Now if ${\cal P}(x)$ is the distribution of $x_{\rm b}$, that gives
the average in Eq. \ref{eq:A8}
\begin{equation}
 \label{eq:A9}
x_{\rm eq}=\langle x\rangle=\int x \,{\cal P}(x)\ dx ,
\end{equation}
then ${\cal P}(x)$ can be written as
\begin{equation}
\label{eq:A10}
{\cal P}(x) = \frac{\int {\cal DX}\ e^{-\beta W}\ \delta (x_{\rm b}-x)}{\int
            {\cal DX} \; e^{-\beta W}} ,
\end{equation}
as quoted in Eq. \ref{eq:3}. We now invoke the moment
theorem\cite{mom} which,  in our case,  states that for a probability
distribution without sufficiently long tails, the moments uniquely
specify the distribution.  Since these conditions are satisfied by
the equilibrium probability distributions for any finite system,
the moment theorem applies.
Since the moments from the nonequilibrium path
integral are the equilibrium moments, ${\cal P}(x)$ is the equilibrium
distribution:~${\cal P}(x)=P_\lambda(x)$. This completes the proof.
\subsection{Generalization}
In general, for a Hamiltonian of the form
$H=H(\{\Lambda_\alpha\},\{X_\alpha\})$, the equilibrium distribution,
$P(E,x_1,x_2,...)$, at some given parameter values,
$\{\lambda_\alpha\}$ and temperature $\beta^{-1}$, can be obtained in
the same way provided the paths start from an equilibrium state for
$H=H_0$, where $H_0$ gives the energy for all $\Lambda_\alpha=0$ and
$W$ is the total work done on the system along a nonequilibrium path,
by each of the externally controlled parameters. $E$ here corresponds
to the energy from $H_0$ only. Our starting $H_0$
may be a free Hamiltonian for a mechanical system and \textit{can as
  well be zero} for interacting  spin-like systems.

Consider the Hamiltonian $H=\gamma H_0$  for a spin-like system
  (i.e. without any kinetic energy). In this case one of the
$\{\Lambda_\alpha\}$ could be the strength of interaction. Let's start
with $\gamma=0$, i.e. the starting point is any random configuration
of the free system or a non-interacting system, and then change
$\gamma$ in some given way from $\gamma=0$ to $\gamma=1$. We thus
generate the equilibrium distribution of $H_0$ at a particular
$\beta$, by doing a similar nonequilibrium path averaging. Note that
everywhere we need the product $\beta W$. So, we can discretize
temperature instead of $\Lambda$ and the process can be reinterpreted
as cooling down to a finite temperature from an initial infinite
temperature. In the usual formulation of work theorem, $\Lambda$
refers to mechanical parameters such as the pulling force in AFM, which are under direct
control of the experimentalists.  In contrast, other intensive
  parameters such as temperature may not be controlled with this level
  of precision in experiments. But this finds various applications in
numerical experiments.  Such thermal quenches are quite common in
numerical simulations and our results show how these can be harnessed
to extract equilibrium information as well.  The ensemble of states
obtained in the above discussed way at the end of the path is {\it
  not} a representative sample of the equilibrium ensemble at the
concerned temperature and field. However, the history-averaged
distribution is the equilibrium distribution. The boundary states
would relax to reach equilibrium via energy transfer to the reservoirs
but that part of the process is not required.  This difference becomes
important and visible in systems exhibiting hysteresis as e.g. for a
ferromagnet.
\subsection{Application to ferromagnet to get equilibrium magnetization curve}
The above-mentioned scheme can be used to get the equilibrium
probability distribution or thermodynamic quantity from a process
which is arbitrarily away from equilibrium and at all temperatures
including phase transition points.  Now we apply our result to the
case of hysteresis of a ferromagnet below the critical temperature
($T_C$). Consider a Hamiltonian: $H=H_0-hM$. The external magnetic
field is varied from $-h_0$ to $+h_0$ in a fixed manner and then
reversed. $\langle M\rangle $ is calculated using Eq. \ref{eq:A8}.
Below the critical temperature, magnetization ($M$) vs. magnetic field
($h$) curve shows a discontinuity at $h=0$ for infinite system size.
For a finite system there is no discontinuity, $M$-$h$ curve is
continuous passing through the origin, and the slope of $M$-$h$ curve
at $h=0$ increases as system size increases. But, in reality, when
experiments or simulations are done, instead of single retraceable
curve passing through the origin we get a loop called hysteresis loop,
no matter how slowly we vary the magnetic field. The common technique
known to get the equilibrium curve is to connect the vertices of the
sub-loops \cite{hyst}. Here the weighted nonequilibrium path integral
scheme is a way out to get the equilibrium magnetization curve. We
verify this for Ising ferromagnet and discuss the observations about
it in Sec. \ref{sec:num}.
\section{Equilibrium probability distribution from an eigenvalue equation: Operator ${\cal S}$}\label{sec:opS}
In this section we derive the main result of this paper: equilibrium probability distribution as an eigenfunction of a nonequilibrium operator ${\cal S}$.

Using the discrete notation, we can write Eq. \ref{eq:3} as
\beq
\label{eq:6}
P_{{\lambda}}(x) = \frac{Z_{\lambda_0}}{Z_{\lambda}}\sum_{\rm paths}\, e^{-\beta W}\delta_{x_{\rm b},x}\ ,
\eeq
by using the work theorem, Eq. \ref{eq:2}, that
\beq
\label{eq:28}
\sum_{\rm paths}e^{-\beta W} =\frac{Z_{\lambda}}{Z_{\lambda_0}}.
\eeq
Again, writing $\sum_{\rm paths}=\sum_{x_{\rm i}}P_{\lambda_0}(x_{\rm i})\sum^{'}_{\rm paths}$, where the primed summation denotes the sum for fixed initial value of $x=x_{\rm i}$
with appropriate probability and $P_{\lambda_0}(x_{\rm i})$ denotes
the equilibrium distribution of $x_{\rm i}$ for $\Lambda={\lambda}_0$,
we get,
\beq
\label{eq:7}
P_{\lambda}(x)=\frac{Z_{\lambda_0}}{Z_{\lambda}}\sum_{x_{\rm i}}\sum_{\rm paths}{}^{^{\prime}} P_{\lambda_0}(x_{\rm i})\, e^{-\beta W}\ \delta_{x_{\rm b},x}.
\eeq
Use the transformation rule for the partition function (Sec. \ref{sec:histo}),
\beq
\label{eq:8}
\frac{Z_\lambda}{Z_{\lambda_0}}=\sum_x P_{{\lambda}_0}(x)\ e^{\beta
     ({\lambda}-{\lambda}_0)x} ,
\eeq
to absorb $Z_{\lambda_0}/Z_{\lambda}$ into the probability
distribution. This transforms $P_{\lambda_0}(x_{\rm i})$ into
$P_{\lambda}(x_{\rm i})$, in Eq. \ref{eq:7} as
\bea
P_{\lambda}(x)&=&\sum_{x_{\rm i}}\sum_{\rm paths}{}^{^{\prime}}\,e^{-\beta W-\beta
({\lambda}-{\lambda}_0)x_{\rm i} }\,\delta_{x_{\rm b},x} P_{\lambda}(x_{\rm i}) \label{eq:9}\\
&=&\sum_{x_{\rm i}} {\cal S}_{x,x_{\rm i}}\, P_{\lambda}(x_{\rm i})\label{eq:10} .\\
&\Rightarrow& {\cal S}\,{\mathbb P}_{\lambda}={\mathbb
  P}_{\lambda}\label{eq:11} ,
\eea
with ${\mathbb P}_{\lambda}$ as a column vector of $\{P_\lambda(x)\}$
and the matrix elements of ${\cal S}$ as
\beq
\label{eq:12} {\cal S}_{x_{\rm f},x_{\rm i}} = \sum_{\rm
  paths}{}^{^\prime} \ e^{-\beta W-\beta ({\lambda}-{\lambda}_0)x_{\rm
    i}}.
\eeq
The summation in Eq. \ref{eq:12} is over all paths that start from an
equilibrium distribution of $\Lambda={\lambda}_0$ with value of $x$ as
$x_{\rm i}$ and end in a state with ${\Lambda}=\lambda$ and $x=x_{\rm
  f}$, with proper normalization (denoted by prime).

Although we use the simple Hamiltonian: $H=H_0-\Lambda\, x$ in the
construction,  Eq.\ref{eq:12} can be generalized for a Hamiltonian
$H=H+H_1(\Lambda,x)$, because  Eq. \ref{eq:8}  has the general form,
$$\frac{Z_\lambda}{Z_{\lambda_0}}=\sum_x P_{{\lambda}_0}(x)\ e^{-\beta\,
     \left[ H(\lambda, x)-H(\lambda_0, x)\right] }. $$

Now we address the remaining problem -- the normalization of the
primed summation over paths in Eq. \ref{eq:12}. This problem is
inherited from Eq. \ref{eq:28}. Note that the l.h.s. of
Eq. \ref{eq:28} should add up to $1$  for
$\lambda=\lambda_0$  with $W=0$. So we choose the hidden factor
{\it a posteriori} by
demanding proper normalization of
the final probability distribution.
This condition can be ensured in a process- or system-independent
way by choosing
$\sum_x {\cal S}_{x,x_{\rm i}}=f(x_{\rm  i})=1$,
(Eq.\ref{eq:10}), i.e. by making the column sum of ${\cal  S}$
independent of
$x_{\rm i}$.  By this normalization of the sum of each column to unity
it is also guaranteed that the principal eigenvalue is $1$. The
corresponding right principal eigenvector has all the elements real
and non-negative -- a necessary condition to be a probability
distribution and when normalized, such that sum of all elements is
unity, this eigenvector gives the equilibrium probability distribution.

The number of rows and columns in ${\cal S}$ is determined by the
number of allowed values of $x$. For continuum of states, the matrix
equation is to be replaced by an integral eigenvalue equation.

Hence, in brief, the scheme to get the equilibrium distribution at
some parameter value ${\lambda}$ and temperature $\beta^{-1}$ is as
follows: Pre-fix some arbitrary or convenient-to-start-with initial
parameter value ${\lambda}_0$ which will be same for all
paths/experiments. Choose a microstate from the equilibrium
distribution at field ${\lambda}_0$ and call its value of $x$ as
$x_{\rm i}$. Change the parameter value from ${\lambda}_0$ to
${\lambda}$ in some predetermined way and measure the work done by the
external parameter on the system according to Eq. \ref{eq:1}. Repeat
the experiments several times and construct the matrix ${\cal S}$
using Eq. \ref{eq:12}. Next, each column of the matrix is normalized
to unity. The normalized principal eigen-vector is the equilibrium
probability distribution, $P_{\lambda}(x)$, at the field ${\lambda}$.

Eq. \ref{eq:11} is the main result of this paper and it is not
restricted to one external parameter only and can be generalized to
any parameter as mentioned above. The matrix ${\cal S}$ connects any
two allowed states of the system without any reference to equilibrium
anywhere and yet its principal eigen-vector determines the equilibrium
distribution. Despite resemblance, there is no similarity either with
the stochastic matrix of a Markov process or the adiabatic switching
on of interaction in a quantum system because $\cal S$ is constructed
out of a finite process and needs global information about the work
done.

Another issue that comes up in this approach via ${\cal S}$, is the
question of ergodicity which connects the Gibbsian statistical
mechanics with equilibrium thermodynamics.
The nonequilibrium dynamics used to construct ${\cal S}$
may not respect ergodicity but the starting points for the paths in
principle span the whole phase space, even in the case when one starts
with a free non-interacting system.  It seems ergodicity of the free
noninteracting system is sufficient to generate the equilibrium distribution.
\subsection{Examples}
\subsubsection{Example 1: Extreme cases}

Consider an extreme case: a completely equilibrium evolution of the
system, where at each step the system reaches its equilibrium. Take a
simple system: a single spin problem in magnetic field $h$ and
temperature $\beta^{-1}$: $\beta H=-Ks$, where $s=\pm 1$ and $K=\beta
h$. For an $n$-step process, $K$ varies from $0$ to $nk$ in steps of
$k$, and the column normalized ${\cal S}$ matrix can be calculated
exactly where at each step the spin reaches the corresponding
equilibrium state, as
\beq
\label{eq:13}
\cal S=\left(
\begin{array}{cc}
P_{nk}(+) & P_{nk}(+)\\
P_{nk}(-) & P_{nk}(-)
\end{array}
\right) ,
\eeq
where $P_{nk}(\pm)$ is the equilibrium probability of
finding $\pm 1$ spin at the $n$-th step. Thus for a completely
equilibrium evolution of the system the elements of the matrix ${\cal
  S}$ are unique and, therefore, ${\cal S}$ has only one and unique
eigenvector. In that case principal eigenvalue is $1$ and all other
eigenvalues are zero. We may conclude that a complete reducibility of
${\cal S}$ is the signature of a thermodynamic process.

Eq. \ref{eq:13} is to be compared with the extreme nonequilibrium
process as embodied in Eq. \ref{eq:5}. For this instantaneous change
in $\lambda$, ${\cal S}=\mathbb {I}$, the identity matrix, with no zero eigenvalues.

If at each of these $n$ steps, the system evolves for a time $\Delta t$
in contact with the bath, then ${\cal S}_{n,\Delta t}\rightarrow {\cal
  S}_{eq}$ as $\Delta t\rightarrow \infty$. The smallness of the rest
of the eigenvalues would indicate how close to equilibrium the system
is.

The dynamics of a many body system might be compartmentalized into
slow modes and fast modes, where the fast modes would equilibrate much
more quickly than slow ones. How many such fast modes have actually
equilibrated, can be gauged by the number of zero eigenvalues. The
${\cal S}$ matrix is not necessarily symmetric, though real and there
is a possibility of pairs of complex conjugate eigenvalues, with their
magnitudes going to zero as equilibrium is reached.
\subsubsection{Example 2: Barkhausen noise and matrix ${\cal S}$}
We now show the practical feasibility of the operator method for a
magnet by using the Barkhausen noise\cite{bar,hyst} 
as recorded through
the output voltage across a secondary coil wound around a
ferromagnetic material.
Though Barkhausen noise has seen many applications, its use for equilibrium properties has not been anticipated.

Consider the Hamiltonian
\beq
\label{eq:14}
H=H_0-h\,M .
\eeq
Here magnetic field $h$ and magnetization $M$ correspond to $\Lambda$
and $x$ respectively. The field is varied from $h_{\rm i}$ to $h_{\rm
  f}$ in a time interval $\tau$ at a constant rate $\dot h$. The
Barkhausen effect is a noisy signal proportional to the change in
magnetization, $\eta(t)= \frac{dM(t)}{dt}$. So by integrating the
Barkhausen noise up to time $t$ one gets the nonequilibrium
instantaneous magnetization of the material. Therefore, we can write
the work related exponent in Eq. \ref{eq:12} as
\beq
\label{eq:15}
W+\left[ h(\tau)-h(0)\right]M_{\rm i}=-\dot{h}\int_0^\tau dt\int_0^{t}
\eta(t')\ dt',
\eeq
which, in a discretized form, looks like
\beq
\label{eq:16}
W+\left[ h_{\rm f}-h_{\rm i}\right]M_{\rm i}=-\Dh\sum_{j=1}^{n-1}
\sum_{k=1}^{j} \eta_k ,
\eeq
where the Barkhausen noise at $k$-th step is $\eta_k=M_k-M_{k-1}$.
Hence the matrix elements ${\cal S}_{M_{\rm f},M_{\rm i}}$ takes the
form
\beq
\label{eq:17}
{\cal S}_{M_{\rm f},M_{\rm i}}=\sum_{expts.}{}^{^\prime} \exp \left[ {\beta \Dh\sum_{j=1}^{n-1} \sum_{k=1}^{j} \eta_k}\right] ,
\eeq
expressed entirely in terms of the Barkhausen noise along the
nonequilibrium paths. The primed summation over paths that start with
$M_{\rm i}$ and end at $M_{\rm f}$ includes proper normalization as
mentioned earlier.

To go to other cases, e.g., for the case of a polymer pulled at a
constant rate of change of force, one needs to monitor the time
variation of the pulled point displacement $dx/dt$ vs
$t$.  This information can then be used in Eq. \ref{eq:17} to get the corresponding ${\cal S}$.
\section{Numerical verification of results}\label{sec:num}
Our claims about the probability have been verified for the case of
$2D$ Ising model on a square lattice, $L\times L$, where $L$ is the
size of the lattice with periodic boundary condition. Consider the
Hamiltonian
\beq
\label{eq:18}
H=-J\sum_{<k,l>}s_k s_l-h\sum_k s_k , \eeq where $J$ is the
interaction strength, $h$ is the external magnetic field and $s_k=\pm
1$ is the spin at $k$-th site of a square lattice. Here
$\sum_{<k,l>}$ denotes the sum over nearest neighbor spins. Here $J$
and $h$ play the roles of external parameter ($\Lambda$) and
$\sum_{<k,l>}s_k s_l$ and $\sum_k s_k$ are the internal variables
($x$).

We find equilibrium probability distribution for given $J$ and $h$
using weighted nonequilibrium path integral, normalizing the
eigenfunction of $\cal S$ and compare those with the equilibrium
probability distribution obtained from a usual Monte Carlo procedure.
The overlap of the two distributions is determined by the
Bhattacharyya coefficient\cite{bha} defined as
\beq
 \label{eq:19}
BC=\sum_{E,M}\sqrt{P_h(E,M)P_{eq}(E,M)}=1-\epsilon ,
\eeq
with $BC=0$ for no overlap and $BC=1$ for complete overlap.
\subsection{Numerical verification of the equilibrium probability
  distribution starting from a uniform distribution}
\figtwo
Let us take an $8\times 8$ lattice and start from $H=0$. Each time we start
from a state chosen from a uniform distribution and reach the final
state with $J=1$ and $h=1$ in $n$-steps. At each $i$-th step, $J$ is switched from $J_i$ to $J_{i+1}$ and the external magnetic field
from $h_i$ to $h_{i+1}$, $$\Delta J=J_{i+1}-J_i=J/n\ \ {\rm and}\ \ \Delta
h_i=h_{i+1}-h_i=h/n;$$ keeping the spin configuration unchanged, and
the amount of work done on the system $$W_i=-\Delta J_i E_i-\Delta
h_i M_i,$$ is calculated where $M_i$ is the magnetization and $E_i$ is $\sum s_k s_l$
at the $i$-th step. Then we let the system relax at that field $h_i$,
$J_i$ and $\beta$ for a while, but do not equilibrate. Thus the work
along a path consisting of $n$ steps is $$W=-\sum_{i=0}^{n-1}\Delta J_i
E_i+ \Delta h_iM_i,$$ which is different for different paths.
We find
the weighted distribution
\beq
 \label{eq:20}
 P_{J,h}(E,M)=\frac{\int {\cal DX}\ e^{-\beta W}\delta(E_{\rm
     b}-E)\delta(M_{\rm b}-M)}{\int {\cal DX}\ e^{-\beta W}},
\eeq
and then $$P_{J,h}(M)=\sum_E P(E,M)$$ and $$P_{J,h}(E)=\sum_M P(E,M).$$ It is observed
that these distributions merge well with the corresponding equilibrium
distributions and for $P_{J,h}(E)$ (Fig.\ref{fig:fig2}(a)) and $P_{J,h}(M)$
(Fig.\ref{fig:fig2}(b)) we get $\epsilon\sim 10^{-3}$ (Eq. \ref{eq:19}).

\subsection{Equilibrium magnetization curve using nonequilibrium path integral}
For this case lattice size is $8\times 8$ and the interaction strength is kept fixed
at $J=1$. Each time we start from an equilibrium distribution of
$h=-h_0$. The field is varied from $-h_0$ to $+h_0$ in $n$ steps.
$W(n)$ vs. $n$ data are recorded and $\langle M\rangle(h)$ is
calculated using Eq. \ref{eq:A8}.

We plot the weight averaged magnetization curve, $\langle
M\rangle(h)$, along with the hysteresis loop, average magnetization
over samples, against $h$ for $h_0=0.2$ in Fig.\ref{fig:fig3} and
$h_0=2$ in Fig. \ref{fig:fig4}.
\figthree

A retraceable equilibrium curve is obtained as expected though the
nominally averaged magnetization neither changes sign nor makes a
complete loop (Fig.\ref{fig:fig3})\cite{bkc}.  This reflects the fact
that though in majority the magnetization does not reach the correct
value, there are a few rare samples for which the spins do flip and
these rare configurations, which are close to equilibrium, get more
weight in the weighted path integral to give the correct equilibrium
curve.
\figfour

For the larger field, we obtain a curve which is much narrower
than the hysteresis curve (Fig.\ref{fig:fig4}). The equilibrium curve obtained
this way is still not a
single curve. The width of the loop might be connected
to the droplet time scale, and signals the need for a more careful sum
over paths to take care of droplet fluctuations.
%
%
%
\subsection{Numerical verification of the eigenvalue equation}

We start from an equilibrium ensemble at inverse temperature
$\beta=0.2$ (kept fixed throughout the experiment), $J=1$ and $h=0$.
Each time we start from a state chosen from its equilibrium
distribution and reach the final state with $J=1$ and $h=1$ in
$n$-steps in the same way described above and calculate the amount of
work on the system at $i$-th step: $W_i=-\Delta h_i M_i$. We find the
matrix elements:
\beq
\label{eq:21} {\cal S}_{M_{\rm f},M_{\rm i}} = \sum_{\rm
  paths}{}^{^{\prime}} \ e^{-\beta W-\beta (h-h_0)M_{\rm
    i}}\,\delta_{M_{\rm b},M_{\rm f}} .
\eeq
After the matrix is constructed, we normalize sum of each column to unity and find the normalized principal eigen-vector corresponding to the
Principal eigenvalue $1$, which is guaranteed. We compare the normalized eigenfunction with the actual equilibrium distribution for
$L=4\ {\rm and}\ 8$. We see that these distributions merge with the
corresponding equilibrium distributions for $L=4$ (Fig.\ref{fig:fig5}(a)) and $L=8$ (Fig.\ref{fig:fig5}(b)) with $\epsilon\sim 10^{-4}$ (Eq. \ref{eq:19}).
\figfive
\section{Summary}\label{sec:summ}

In this paper we show and verify numerically that the
repeated nonequilibrium measurements of work done to connect any two
microstates of a system can be used to construct a matrix ${\cal S}$
whose principal eigenvector is the equilibrium distribution.
The matrix elements of ${\cal S}$ (Eq. \ref{eq:12}) for a Hamiltonian $H(\Lambda,x)$ with $(\Lambda,x)$ as a conjugate pair are:
\beq
{\cal S}_{x_{\rm f},x_{\rm i}} = \sum_{\rm
  paths}{}^{^\prime} \ e^{-\beta W+\beta \left[ H(\lambda,x_{\rm i})-H(\lambda_0,x_{\rm
    i})\right] }
\eeq
where the summation is over all paths that start from an equilibrium distribution
of externally controlled parameter $\Lambda = \lambda_0$ with value of conjugate variable $x$ as $x_i$ and end in a state with $\Lambda = \lambda$ and $x = x_f$, with proper
normalization. The work done $W$ is defined in Eq. \ref{eq:1}.
The values of the elements of ${\cal S}$ depend on the details of the
process and, therefore, there can be many different ${\cal S}$, but
all will have the same invariant principal eigenvector. In this way
the distribution of an interacting system can be obtained from a free,
non-interacting one without any reference to equilibrium anywhere.
In the process, we also provide a dynamics independent proof of the
result that the equilibrium probability distribution
can be obtained using the nonequilibrium path integral.
Besides giving a new perspective of thermodynamics and statistical
mechanics, our result has direct implications for new ways in
numerical simulations and experiments.

\end{document}